\documentclass[runningheads]{llncs}
\usepackage{algorithm}
\usepackage{anyfontsize}
\usepackage{algpseudocode}
\usepackage[T1]{fontenc}
\usepackage{amsfonts}
\usepackage{physics}
  
\usepackage{amsmath}
\usepackage{amsthm}
\usepackage{booktabs}
\usepackage{color}
\usepackage{xcolor}
\usepackage{colortbl}
\usepackage{esvect}
\usepackage{graphicx}
\usepackage{hyperref}
\usepackage[capitalise]{cleveref}
\usepackage{lineno}
\usepackage{longtable}
\usepackage{makecell}
\usepackage{multirow}
\usepackage{caption}
\usepackage{subcaption}
\usepackage{stmaryrd}
\usepackage{tabularx}
\usepackage{tikz}
\usepackage{tikz-network}
\usepackage{xspace}
\usepackage{xfrac}
\usepackage{pgfplots}
\usepackage{svg}
\usepackage{microtype}
\usepackage{nicefrac}
\usepackage{siunitx}
\usepackage{numprint} 
\usetikzlibrary{positioning,calc}
\usetikzlibrary{positioning,arrows,petri,shapes,fit,automata}
\usepgfplotslibrary{fillbetween}
\usepackage{filecontents}
\def\lmodel{kNN\xspace}
\def\lmodel{SLPN\xspace}

\theoremstyle{definition}
\newcommand{\myparagraph}[1]{\smallskip\noindent\textbf{#1}}
\usepackage[figurename=Fig.]{caption}
\DeclareMathOperator{\multisetminus}{\mathbin{{\setminus}\mspace{-7mu}{-}}}
\tikzset{fontscale/.style = {font=small}}
\def\lmodel{SLPN\xspace}
\begin{document}
\title{
Stochastic Alignments: Matching an Observed Trace to Stochastic Process Models
}
\author{Tian Li\inst{1,2}{\href{https://orcid.org/0000-0003-1288-3149}{\protect\includegraphics[scale=0.05]{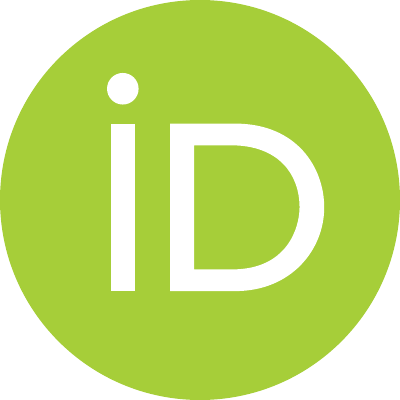}}}\and
Artem Polyvyanyy\inst{2}{\href{https://orcid.org/0000-0002-7672-1643}{\protect\includegraphics[scale=0.05]{orcid}}}
\and
Sander J.J. Leemans\inst{1,3}{\href{https://orcid.org/0000-0002-5201-7125}{\protect\includegraphics[scale=0.05]{orcid}}}}
\authorrunning{Tian Li \and Artem Polyvyanyy \and Sander J.J. Leemans}
\institute{RWTH Aachen University, Germany \\
\email{t.li,s.leemans@bpm.rwth-aachen.de}\and
The University of Melbourne, Australia\\
\email{artem.polyvyanyy@unimelb.edu.au}\and Fraunhofer, Germany}
\titlerunning{Stochastic Alignments }
\maketitle              
\begin{abstract}
Process mining leverages event data extracted from IT systems to generate insights into the business processes of organizations.
Such insights benefit from explicitly considering the frequency of behavior in business processes, which is captured by stochastic process models.
Given an observed trace and a stochastic process model, conventional alignment-based conformance checking techniques face a fundamental limitation: They prioritize matching the trace to a model path with minimal deviations, which may, however, lead to selecting an unlikely path.
In this paper, we study the problem of matching an observed trace to a stochastic process model by identifying a likely model path with a low edit distance to the trace.
We phrase this as an optimization problem and develop a heuristic-guided path-finding algorithm to solve it.
Our open-source implementation demonstrates the feasibility of the approach and shows that it can provide new, useful diagnostic insights for analysts.
\end{abstract}

\section{Introduction}
A process is a series of coordinated steps.
In modern organizations, information systems record processes executed by employees, managers, and customers as event data.
These data can be extracted into event logs comprising sequences of executed events, called \emph{traces}, triggered by different historical process instances.
Organizations typically establish detailed guidelines for process execution to help achieve their objectives~\cite{DBLP:books/wi/Dumas2005}.
These guidelines are often represented as \emph{process models} that reflect standards or regulations~\cite{DBLP:books/sp/Aalst16} for executing the processes.

Process mining is a data analysis technique that aims to deliver meaningful insights from event data.
In particular, techniques that explicitly consider the frequency or likelihood of process behavior are categorized as stochastic process mining approaches.
For instance, to avoid the risk that effort is spent on rare behavior and leads to misinformed decisions, it is vital to distinguish between frequent and rare process behavior~\cite{DBLP:journals/is/LeemansP23}. 
Stochastic process mining techniques achieve this by treating an event log as a finite sample drawn from a probability distribution over traces, while modeling stochastic processes as probability distributions over traces~\cite{Alkhammash2024}.

Conformance checking techniques such as alignments enable analysts to find
commonalities and discrepancies between the modeled behavior and the observed
behavior~\cite{DBLP:books/sp/22/0001DW22}. 
Conventional alignments-based techniques~\cite{Adriansyah2014AligningOA,DBLP:conf/bpm/Dongen18} identify a path allowed by the model with as few deviations as possible from an observed trace.
However, when considering a stochastic perspective, if the selected path is unlikely according to the stochastic process model, it may not be the most likely explanation of the path through the model.
As a consequence, further diagnostics are based on process behavior that is less relevant to be followed by design.

For instance, a path with a probability of $10\%$ according to the model and an edit distance of 3 to the trace may be a better match than a path with a probability of $0.1\%$ and an edit distance of 2.
The scenario highlights two possibly competing objectives when matching the trace to the stochastic model of the process: the probability of the selected path allowed by the model and its edit distance to the trace.

In this paper, we propose a \emph{stochastic alignment} technique, which matches a single trace to a stochastic process model and produces an alignment that balances the importance of the normative behavior (the probability of the model path) and alignment cost (the edit distance between model path and trace).
Business analysts can explicitly weigh the trade-off with a user-defined parameter.
For a specific type of stochastic model, stochastic labeled Petri net, we formulate the search for the model path as an optimization problem solved by the a-star algorithm~\cite{DBLP:journals/tssc/HartNR68}.
Our technique has been implemented and is publicly available.
The experiments demonstrate that the technique is feasible for real-life event data.
Moreover, the case study shows that it can generate new and useful insights when explaining deviations in observed traces.

The remainder of the paper proceeds as follows. We first introduce preliminaries in~\cref{sec:preliminaries}. In~\cref{sec:method}, we present our technique and then evaluate it in~\cref{sec:evaluation}. Then, we discuss related work in~\cref{sec:related_work}. Finally, \cref{sec:conclusion} draws conclusions and sketches future work.

\section{Preliminaries\label{sec:preliminaries}}
In this section, we present concepts used in the subsequent sections.

Given a set of elements $E$, a multiset $X: E\rightarrow \mathbb{N}$ maps the elements of $E$ to the natural numbers, such that $X$ allows for multiple instances for each element.
Multisets are shown within $[\ ]$ with superscript frequencies. 
For example, $X = [b^4,c^5]$ is a multiset with four $b$'s and five $c$'s. 
Multiset union $X_u = X_1\uplus X_2$ denotes that $\forall_{e\in E}\ X_u(e) = \max(X_2(e), X_1(e))$. 
Multiset subset $X_1\subsetpluseq X_2$ denotes $\forall_{e\in E}X_2(e)\ge X_1(e)$. 
If $X_1\subsetpluseq X_2$, then $X_3=X_2\multisetminus X_1$ is the multiset difference, such that $\forall_{e\in E}X_3(e)=X_2(e)-X_1(e)$.
The set of all multisets on $E$ is denoted by $\mathcal{B}(E)$.
Given an alphabet $\Sigma= \{a_1,\ldots,a_n\}$, the Parikh vector of a sequence over $\Sigma$ is a function $\vv{\cdot}\colon \Sigma^* \rightarrow \mathbb{N}^{|\Sigma|}$ that results in a column vector such that $\vv{\sigma}=[|\sigma|_{a_1},\ldots,|\sigma|_{a_n}]$, where $|\sigma|_{a_i}$ denotes the number of $a_i$ in $\sigma$. 

An \emph{event log} is a collection of \emph{traces}.
Each trace is a finite sequence of \emph{activities}. 
An activity is a description of the event that is occurring.
The control-flow perspective of processes can be captured using labeled Petri nets.

\begin{definition}[Labeled Petri Nets]
A \emph{labeled Petri net} (LPN) is a tuple $(P, T, F, A, \rho, M_0)$ where $P$ is a finite set of places, $T$ is a finite set of transitions such that $P \cap T = \emptyset$, $F \subsetpluseq \mathcal{B}((P \times T) \cup (T \times P))$ is a flow relation, $A$ is a finite set of activities, $\rho \colon T \rightarrow A \cup \{\tau\}$ with $\tau \not\in A$ is a labeling function, and $M_0 \in \mathcal{B}(P)$ is the initial marking.
\end{definition}

We adopt the dot notation to refer to components in tuples.
For example, given an LPN $N$, its transitions are indicated by $N.T$. 
A transition $t$ with $\rho(t) = \tau$ is \emph{silent}. 
${}^{\bullet}t = \{p \in P \mid (p, t) \subsetpluseq F\}$ and $t^{\bullet} = \{p \in P \mid (t, p) \subsetpluseq F\}$ denote the \emph{preset} and \emph{post-set} of $t$.
A state in LPN is called \emph{marking} $M$ that marks certain places (represented by circles) with tokens (represented by black dots), such that $M \in \mathcal{B}(P)$. 
By $T_M$, we denote the set of all transitions enabled at marking $M$, that is, $T_M = \{ t \in T \mid {}^{\bullet}t \subsetpluseq M\}$. 
A \emph{deadlock} marking does not enable any transition.
Firing an enabled transition $t \in T_M$ results in a new marking $M^\prime = (M \multisetminus {}^{\bullet}t ) \uplus t^{\bullet}$.
A \emph{model path} is a sequence of transitions $\eta=\langle t_1, \ldots, t_n\rangle$ that brings the LPN from its $M_0$ to a deadlock marking, that is, there is a sequence of markings $\langle M_0, \ldots ,M_{n} \rangle$ for which it holds that $\forall_{ 1 \leq i \leq n} {}^{\bullet}t_i \subsetpluseq M_{i-1} \land M_i = (M_{i-1}  \multisetminus {}^{\bullet}t_i)\uplus t_i^{\bullet}$ and $M_n$ is a deadlock marking. 
The projection of a model path by $\rho$ on the labels of its non-silent transitions is a \emph{trace}.

For instance, \cref{fig:example_lpn} illustrates an LPN $N_1$.
The initial marking of $N_1$ is $[p_1^1]$, and there are two reachable deadlock markings, i.e., $[p_4^1]$ and $[p_4^2]$.
$\langle t_1,t_3\rangle$ is a model path that brings $N_1$ from $[p_1^1]$ to $[p_4^1]$, and it can be projected to trace $\langle a,c \rangle$.
$\langle t_2,t_3,t_4\rangle$ is another model path that brings $N_1$ from $[p_1^1]$ to $[p_4^2]$, and it can be projected to trace $\langle b,c,d \rangle$.

\begin{figure}[t]
  \begin{minipage}[b]{0.45\textwidth}
    \centering
            \begin{tikzpicture}
  \node[place,tokens=1,minimum width = 0.3cm,minimum height=0.35cm] (p1) [] {};
  \node[transition, minimum width = 0.3cm,minimum height=0.4cm] (t1) [above right=0.5cm of p1] {$a$};
   \node[transition, minimum width = 0.3cm,minimum height=0.4cm] (t2) [below right=0.5cm of p1] {$b$};
  \node[place,minimum width = 0.3cm,minimum height=0.35cm] (p2) [right=0.4cm of t1] {};     \node[place,minimum width = 0.3cm,minimum height=0.35cm] (p3) [right=0.4cm of t2] {};
   \node[transition, minimum width = 0.3cm,minimum height=0.4cm] (t3) [right=0.5cm of p2] {$c$};  
   \node[transition, minimum width = 0.3cm,minimum height=0.4cm] (t4) [right=0.5cm of p3] {$d$};   
   \node[place,minimum width = 0.3cm,minimum height=0.35cm] (p4) [below right=0.4cm of t3] {}; 
  \draw[->] (p1) -- (t1);
  \draw[->] (p1) -- (t2);
  \draw[->] (p2) -- (t3);
  \draw[->] (p3) -- (t4);
  \draw[->] (t1) -- (p2);
  \draw[->] (t2) -- (p2);
  \draw[->] (t2) -- (p3);
  \draw[->] (t3) -- (p4);
  \draw[->] (t4) -- (p4); 
  \node[left=0.06cm of t1,align=center] {\small $t_1$};
  \node[above=0.06cm of t2,align=center] {\small $t_2$};
  \node[right=0.06cm of t3,align=center] {\small $t_3$};
  \node[above=0.06cm of t4,align=center] {\small $t_4$};
  \node[below=0.06cm of p1,align=center] {\small $p_1$};
  \node[below=0.06cm of p2,align=center] {\small $p_3$};  
  \node[below=0.06cm of p3,align=center] {\small $p_2$};
  \node[below=0.06cm of p4,align=center] {\small $p_4$};    
\end{tikzpicture}    
    \caption{Labeled Petri net $N_1$.}
    \label{fig:example_lpn}
  \end{minipage}
  \hfill
  \begin{minipage}[b]{0.45\textwidth}
    \centering
           \begin{tikzpicture}
  \node[place,tokens=1,minimum width = 0.3cm,minimum height=0.35cm] (p1) [] {};
  \node[transition, minimum width = 0.3cm,minimum height=0.4cm] (t1) [above right=0.5cm of p1] {$a$};
   \node[transition, minimum width = 0.3cm,minimum height=0.4cm] (t2) [below right=0.5cm of p1] {$b$};
  \node[place,minimum width = 0.3cm,minimum height=0.35cm] (p2) [right=0.4cm of t1] {};     \node[place,minimum width = 0.3cm,minimum height=0.35cm] (p3) [right=0.4cm of t2] {};
   \node[transition, minimum width = 0.3cm,minimum height=0.4cm] (t3) [right=0.5cm of p2] {$c$};  
   \node[transition, minimum width = 0.3cm,minimum height=0.4cm] (t4) [right=0.5cm of p3] {$d$};   
   \node[place,minimum width = 0.3cm,minimum height=0.35cm] (p4) [below right=0.4cm of t3] {}; 
  \draw[->] (p1) -- (t1);
  \draw[->] (p1) -- (t2);
  \draw[->] (p2) -- (t3);
  \draw[->] (p3) -- (t4);
  \draw[->] (t1) -- (p2);
  \draw[->] (t2) -- (p2);
  \draw[->] (t2) -- (p3);
  \draw[->] (t3) -- (p4);
  \draw[->] (t4) -- (p4); 
  \node[left=0.01cm of t1,align=center] {\small $t_1$};
  \node[above=0.06cm of t2,align=center] {\small $t_2$};
  \node[right=0.01cm of t3,align=center] {\small $t_3$};
  \node[above=0.06cm of t4,align=center] {\small $t_4$};
  \node[below=0.06cm of t1,align=center] {\small $1$};
  \node[below=0.06cm of t2,align=center] {\small $99$};
  \node[below=0.06cm of t3,align=center] {\small $3$};
  \node[below=0.06cm of t4,align=center] {\small $2$};
  \node[below=0.06cm of p1,align=center] {\small $p_1$};
  \node[below=0.06cm of p2,align=center] {\small $p_3$};  
  \node[below=0.06cm of p3,align=center] {\small $p_2$};
  \node[below=0.06cm of p4,align=center] {\small $p_4$};    
\end{tikzpicture}    
    \caption{Stochastic labeled Petri net $N_2$.}
    \label{fig:example_slpn}
  \end{minipage}
\end{figure}

The trace net of a trace is an LPN without choices, such that each place and transition has at most one incoming and at most one outgoing arc.
In~\cref{eq:consume}, \emph{consumption matrix} $\mathcal{C}_N$ of $\abs{N.P}$ rows and $\abs{N.T}$ columns specifies the necessary tokens for firing transitions in LPN $N$.
In~\cref{eq:incidence}, \emph{incidence matrix} $\mathcal{I}_N$ of $\abs{N.P}$ rows and $\abs{N.T}$ columns reflects the flow relation in $N$.
\begin{equation}\label{eq:consume}
\mathcal{C}_N(p,t)=\left\{
             \begin{array}{rl}
              -1 & \quad\textit{if (p,t) $\! {\in} \!$ N.F and (t,p) $\! {\not \in}\!$ N.F, and}\\
              0 & \quad\textit{otherwise.}
             \end{array}
\right.
\end{equation}
\begin{equation}\label{eq:incidence}
\mathcal{I}_N(p,t)=\left\{
             \begin{array}{rl}
              1 & \quad\textit{if (t,p) $\! {\in} \!$ N.F and (p,t) $\! {\not \in}\!$ N.F, and}\\
              C_N(p,t) & \quad\textit{otherwise.}
             \end{array}
\right.
\end{equation}
A variant of LPNs with stochastic information is stochastic labeled Petri nets.

\begin{definition}[Stochastic Labeled Petri Nets]
A \emph{stochastic labeled Petri net} (\lmodel) is a tuple $(P, T, F, A, \rho, M_0, w)$, where $(P, T, F, A, \rho, M_0)$ is an LPN, and $w \colon T \rightarrow \mathbb{R}^+$ is a weight function.
\end{definition}

Given an SLPN, a transition $t$ enabled at marking $M$ can fire with probability \smash{$\mathrm{p}(t \mid M)=\nicefrac{w(t)}{\Sigma_{t^{\prime} \in T_M} w(t^{\prime})}$}.
The probability $\mathrm{p}(\eta)$ of observing path $\eta$ is equal to $\prod_{1 \leq i \leq n} \mathrm{p}(t_i \mid M_{i-1})$. 
For example, \cref{fig:example_slpn} depicts an \lmodel $N_2$, in which transitions $t_1$ and $t_2$ are enabled at initial marking $M_0$.
Transition $t_2$ has a weight of 99, and a firing probability of $\nicefrac{99}{(1+99)} = \nicefrac{99}{100}$ at $M_0$.

Using \lmodel and the trace net, we define the synchronous product model as their combination with an additional set of synchronous transitions, while the transitions of the original \lmodel and the trace net are represented by pairing them with the new symbol $\gg$.

\begin{figure}[t]
  \begin{minipage}[b]{0.64\textwidth}
    \centering
        \begin{tikzpicture}
  \node[place,tokens=1,minimum width = 0.3cm,minimum height=0.35cm] (p1) [] {};
   \node[transition, minimum width = 0.3cm,minimum height=0.4cm] (t2) [below right=0.5cm of p1] {$(\gg,b)$};
     \node[transition, minimum width = 0.3cm,minimum height=0.4cm] (t1) [above=0.8cm of t2] {$(\gg,a)$};
  \node[place,minimum width = 0.3cm,minimum height=0.35cm] (p2) [right=0.7cm of t1] {};  
  \node[place,minimum width = 0.3cm,minimum height=0.35cm] (p3) [right=0.7cm of t2] {};
   \node[transition, minimum width = 0.3cm,minimum height=0.4cm] (t3) [right=0.7cm of p2] {$(\gg,c)$};  
   \node[transition, minimum width = 0.3cm,minimum height=0.4cm] (t4) [right=0.7cm of p3] {$(\gg,d)$};   
   \node[place,minimum width = 0.3cm,minimum height=0.35cm] (p4) [right= 4.5cm of p1] {}; 
  \draw[->] (p1) -- (t1);
  \draw[->] (p1) -- (t2);
  \draw[->] (p2) -- (t3);
  \draw[->] (p3) -- (t4);
  \draw[->] (t1) -- (p2);
  \draw[->] (t2) -- (p2);
  \draw[->] (t2) -- (p3);
  \draw[->] (t3) -- (p4);
  \draw[->] (t4) -- (p4); 
  \node[left=-0.01cm of t1,align=center] {\small $(\gg,t_1)$};
  \node[above=-0.01cm of t2,align=center] {\small $(\gg,t_2)$};
  \node[right=-0.01cm of t3,align=center] {\small $(\gg,t_3)$};
  \node[above=-0.01cm of t4,align=center] {\small $(\gg,t_4)$};
  \node[left=-0.01cm of p1,align=center] {\small $p_1$};
  \node[below=-0.01cm of p2,align=center] {\small $p_3$};
  \node[above=-0.01cm of p3,align=center] {\small $p_2$};
  \node[right=-0.01cm of p4,align=center] {\small $p_4$};
  \node[place,tokens=1,minimum width = 0.35cm,minimum height=0.35cm] (p10) [below left=1.6cm of t2] {};    
  \node[transition,fill={rgb:black,1;white,3}, minimum width = 0.4cm,minimum height=0.4cm] (t10) [right=0.3cm of p10] { $(a,\gg)$};
  \node[place,minimum width = 0.35cm,minimum height=0.35cm] (p20) [right=0.5cm of t10] {};
  \node[transition,fill={rgb:black,1;white,3}, minimum width = 0.4cm,minimum height=0.4cm] (t20) [right=0.4cm of p20] { $(d,\gg)$};
  \node[place,minimum width = 0.35cm,minimum height=0.35cm] (p30) [right=0.5cm of t20] {};
  \node[transition, fill={rgb:black,1;white,3},minimum width = 0.4cm,minimum height=0.4cm] (t30) [right=0.4cm of p30] {$(c,\gg)$};
  \node[place,minimum width = 0.35cm,minimum height=0.35cm] (p40) [right=0.3cm of t30] {};
  \draw[->] (p10) -- (t10);
  \draw[->] (p20) -- (t20);
  \draw[->] (p30) -- (t30);
  \draw[->] (t10) -- (p20);
  \draw[->] (t20) -- (p30);
  \draw[->] (t30) -- (p40);
  \draw[->] (p1) -- (t1);
  \draw[->] (p2) -- (t2);
  \node[below=-0.01cm of p10,align=center] {\small $p_5$};
  \node[below=-0.01cm of p20,align=center] {\small $p_6$};
  \node[below=-0.01cm of p30,align=center] {\small $p_7$};
  \node[below=-0.01cm of p40,align=center] {\small $p_8$};
    \node[below=-0.01cm of t10,align=center] {\small $(t_1^\prime,\gg)$};
  \node[below=-0.01cm of t20,align=center] {\small $(t_2^\prime,\gg)$};
  \node[below=-0.01cm of t30,align=center] {\small $(t_3^\prime,\gg)$};
  \node[transition, fill={rgb:black,1;white,8},minimum width = 0.3cm,minimum height=0.4cm] (t100) [below=1cm of p1] {$(a,a)$};
   \node[transition, fill={rgb:black,1;white,8},minimum width = 0.3cm,minimum height=0.4cm] (t200) [right=1.5cm of t100] {$(d,d)$};
     \node[transition, fill={rgb:black,1;white,8},minimum width = 0.3cm,minimum height=0.4cm] (t300) [right=1.4cm of t200] {$(c,c)$};
    \node[right=-0.01cm of t100,align=center] {\small $(t_1^\prime,t_1)$};
  \node[right=-0.01cm of t200,align=center] {\small $(t_2^\prime,t_4)$};
  \node[right=-0.01cm of t300,align=center] {\small $(t_3^\prime,t_3)$};
\draw[->] (p1) -- (t100);
\draw[->] (p10) -- (t100);
\draw[->] (p2) to [bend right=30] (t300);
\draw[->] (p30) -- (t300);
\draw[->] (p3) -- (t200);
\draw[->] (p20) -- (t200);
\draw[->] (p20) -- (t200);
\draw[->] (t100) -- (p20);
\draw[->] (t100) -- (p3);
\draw[->] (t300) -- (p4);
\draw[->] (t200) -- (p30);
\draw[->] (t200) to [bend right=20] (p4);
\draw[->] (t300) -- (p40);
\end{tikzpicture}   
    \caption{The synchronous product net of $\langle a,d,c\rangle$ and $N_2$.}
    \label{fig:sync_net}
  \end{minipage}
  \hfill
\begin{minipage}[b]{.33\textwidth}
{ 
{
    \newcommand{\myline}{\cline{2-7}}
    \newcommand{\noline}[1]{\multicolumn{1}{c}{#1}}
    \begin{tabular}{c|c||c|c|c|c|c|}
        \myline
        $\gamma_1\colon\ $ & $\sigma$ & $a$   & $\gg$   & $d$    & $c$& $\gg$\\
        \myline
        & $\eta_1$ & $\gg$ & $t_2$ & $\gg$ & $t_3$&$t_4$ \\
        \myline
    \end{tabular}
    }
    \newline
\vspace*{0.2cm}
\newline
    {
    \newcommand{\myline}{\cline{2-6}}
    \newcommand{\noline}[1]{\multicolumn{1}{c}{#1}}
    \begin{tabular}{c|c||c|c|c|c|}
        \myline
        $\gamma_2\colon\ $ & $\sigma$ & $\gg$ & $a$ & $d$&$c$  \\
        \myline
                      & $\eta_2$ & $t_2$ & $\gg$ & $t_4$  &$t_3$ \\
        \myline
    \end{tabular}
    }
     \newline
\vspace*{0.2cm}
\newline
    {
    \newcommand{\myline}{\cline{2-5}}
    \newcommand{\noline}[1]{\multicolumn{1}{c}{#1}}
    \begin{tabular}{c|c||c|c|c|}
        \myline
        $\gamma_3\colon\ $ & $\sigma$ &$a$  & $d$  &  $c$\\
        \myline
      & $\eta_3$ & $t_1$ & $\gg$ & $t_3$ \\
        \myline
    \end{tabular}
    }
    \caption{Example alignments for $\langle a,d,c\rangle$ and $N_2$.}
    \label{fig:alignments}
    }
\end{minipage}
\end{figure}

\begin{definition}[Synchronous Product Nets]
Let $N^\sigma=(P^{\sigma}, $ $T^{\sigma},$ $F^{\sigma}, A^{\sigma},$ $\rho^{\sigma},$ $ M_0^{\sigma})$ be the trace net of trace $\sigma$, and $N^s=(P^s,$ $T^s,$ $ F^s,$ $ A^s,$ $ \rho^s,$ $ M_0, $ $w^s)$ be an \lmodel. Their synchronous product net $N^\otimes=(P^{\otimes}, $ $T^{\otimes},$ $F^{\otimes},$ $A^\otimes, $ $\rho^{\otimes}, $ $M_0^{\otimes})$ is a tuple with:\begin{itemize}  
    \item $P^{\otimes}=P^s \cup P^{\sigma}$ is the set of places,
    \item $T^{\otimes}=\{(t^{\sigma},t^s) \in(T^\sigma \cup\{\gg\}) \times(T^s \cup\{\gg\}) \mid t^\sigma \neq \gg \vee t^s \neq \gg \vee \rho^\sigma(t^\sigma)=\rho^s(t^s)\}$ is the set of original transitions and the synchronous transitions,
    \item $F^{\otimes}=\{((t^\sigma, t^s), p) \in T^{\otimes} \times P^{\otimes} \mid(t^s, p^s) \in F^s \vee(t^{\sigma}, p^\sigma) \in F^{\sigma}\} \cup\{(p,(t^\sigma, t^s)) \in P^{\otimes} \times T^{\otimes} \mid(p^s, t^s) \in F^s \vee(p^\sigma, t^{\sigma}) \in F^{\sigma}\}$ is a flow relation,
    \item $A^\otimes=A^\sigma \cup A^s$,
    \item $\rho^{\otimes} \colon T^\otimes \rightarrow A^\sigma \cup A^s \cup \{\tau\}$ is a labeling function, 
    \item $M_0^{\otimes}=M_0^s \uplus M_0^{\sigma}$ is the initial marking.
    \end{itemize}
\end{definition}

For example,~\cref{fig:sync_net} illustrates a synchronous product net for $N_2$ and a trace $\langle a,d,c\rangle$.
We adopt a similar notation as~\cite{Adriansyah2014AligningOA}, which categorize each transition $t \in N^\otimes.T$ as: 1) a non-silent model move $(\gg,t^s)$ with $\rho^{s}(t^s) \ne \tau$; 2) a silent model move $(\gg,t^s)$ with $\rho^{s}(t^s) = \tau$; 3) a trace move $(t^\sigma,\gg)$ with $\rho^{\sigma}(t^\sigma)\ne \gg$; 4) a synchronous move $(t^\sigma,t^s)$ with $\rho^s(t^s) = \rho^\sigma(t^\sigma)\wedge \rho^s(t^s)\ne \gg \wedge \rho^\sigma(t^\sigma)\ne \gg$.

Given a trace and an \lmodel, an alignment is a sequence of moves in their synchronous product net.

\begin{definition}[Alignments]
Let $N^\sigma$ be a trace net of $\sigma$, $N^s$ be an SLPN, and $N^\otimes$ be their synchronous product net. A model path $\gamma$ from the initial marking to a deadlock marking in $N^\otimes$ is an alignment between $\sigma$ and $N^s$.
\end{definition}

~\cref{fig:alignments} illustrates three alignments, $\gamma_1$, $\gamma_2$ and $\gamma_3$.
By stripping the $\gg$ symbol, the upper row can be projected to the trace, while the lower row can be projected to a model path in the \lmodel.
Typically, transitions of type synchronous moves and silent model moves in the synchronous product net are assigned a cost of zero~\cite{Adriansyah2014AligningOA}.
For other types of transition, we assign a cost of one since they represent deletion or insertion.
Thus, $\gamma_1$, $\gamma_2$ and $\gamma_3$ are alignments with a cost of 4, 2, and 1, respectively.
The total move cost is comparable to the Levenshtein Edit Distance~\cite{DBLP:journals/tssc/HartNR68} between the corresponding model path $\eta$ in \lmodel and the trace $\sigma$, denoted as $d(\sigma,\eta)$\footnote{In this paper, we use edit distance and move cost interchangeably.}.

\begin{figure}[t]
    \centering
    \includegraphics[width=\linewidth]{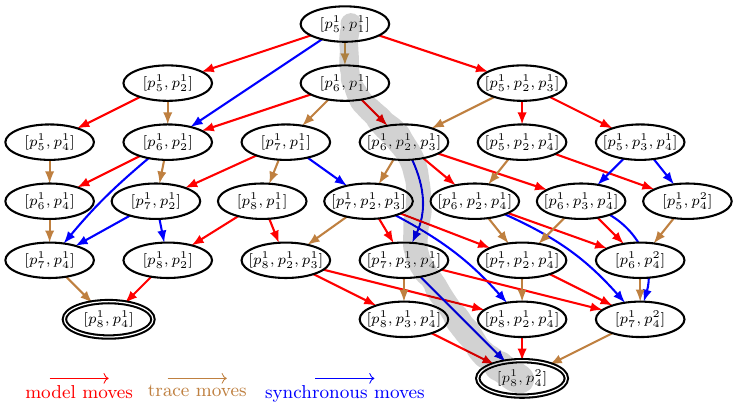}
    \caption{The state space of the synchronous product net in~\cref{fig:sync_net}.}
    \label{fig:state_space}
\end{figure}

\cref{fig:state_space} presents the state space of the synchronous product net in~\cref{fig:sync_net}. 
The highlighted gray path in~\cref{fig:state_space} corresponds to alignment $\gamma_2$, which consists of one trace move $(t_1^\prime,\gg)$, one model move $(\gg,t_2)$, and two synchronous moves $(t_2^\prime, t_4)$ and $(t_3^\prime,t_3)$.

Finally, to establish a direct connection between transitions in the original \lmodel and the model moves and synchronous moves in the synchronous product net, we introduce two reverse functions as follows. 

\begin{definition}[Reverse Functions]
    Let $N^\sigma=(P^{\sigma}, T^{\sigma}, F^{\sigma}, A^{\sigma},\rho^{\sigma}, M_0^{\sigma})$ be the trace net of trace $\sigma$, and $N^s=(P^s, T^s, F^s, A^m, \rho^s, M_0^s, w^s)$ be an \lmodel, and $N^\otimes=(P^{\otimes}, T^{\otimes}, F^{\otimes},A^\otimes, \rho^{\otimes}, M_0^\otimes)$ be their synchronous product net. $r_m\colon\mathcal{B}(P^\otimes) \rightarrow \mathcal{B}(P^s)$ is a function that removes the multiset of places of trace net from marking in $N^\otimes$, and $r_t\colon T^\otimes \backslash{\{\{\gg\}\cup T^\sigma\}} \rightarrow T^s$ is a function that maps the model moves and synchronous moves back to transitions in $N^s$.
\end{definition}

For instance, the initial marking of the synchronous product net $[p_1^1,p_5^1]$ can be mapped to the initial marking $[p_1^1]$ in \lmodel using $r_m$.
The synchronous move $(t_1,t_1^\prime)$ and model move $(t_1,\gg)$ can be mapped to transition $t_1$ in \lmodel with $r_t$.

As the alignment explicitly identifies a model path through the original \lmodel, we can associate a model move or a synchronous move with the probability of the model path.
To do so, we introduce the concept of probability gain based on the reverse functions.

\begin{definition}[Probability Gain]
   Let $N^\otimes=(P^{\otimes}, T^{\otimes},F^{\otimes},A^\otimes, \rho^{\otimes}, M_0^{\otimes})$ 
   be the synchronous product net of an \lmodel $N^s=(P^s$, $T^s,$ $F^s, A^s,\rho^s,$ $M_0^s,$ $w^s)$ and a trace net $N^\sigma$, $r_m$ and $r_t$ be the reverse functions. Then, the probability gain is a function that takes a marking $m^{\otimes}$ of $N^{\otimes}$ and an enabled transition $t^{\otimes} \in T_{m^{\otimes}}$:
   \begin{equation}\label{eq:probability_gain}
\mathrm{p}(m^\otimes,t^\otimes)=\left\{
             \begin{array}{cl} 
              1 & \quad\textit{if }\ t^\otimes \ \textit{ is a trace move}\\
              \frac{w^s(r_t(t^\otimes))}{\Sigma_{t \in T^s_{r_m(m^\otimes)}} w^s(t)} & \quad\textit{otherwise.}
             \end{array}
\right.
\end{equation}
\end{definition}
The probability gain function in~\cref{eq:probability_gain} associates a model move or a synchronous move with the probability of the corresponding model transition, and associates a trace move with probability 1.
For instance, in the synchronous product net shown in~\cref{fig:sync_net}, $(\gg,t_1)$, $(\gg,t_2)$, $(t_1^\prime,\gg)$ and $(t_1^\prime,t_1)$ are enabled at the initial marking $[p_5^1,p_1^1]$.
The probability gain of trace move $(t_1^\prime,\gg)$ is 1.
For $(\gg,t_1)$, $(\gg,t_2)$, and $(t_1^\prime,t_1)$, their probability gains are $\nicefrac{1}{100},\nicefrac{99}{100}$ and $\nicefrac{1}{100}$, respectively.

The probability gain of an alignment is computed as the multiplication of the probability gain of its sequence of moves, which corresponds to the probability of the underlying model path from \lmodel. 
Although we assign a probability gain of 1 to trace moves, this does not introduce bias favoring moves on the trace over moves on the model (whose probability depends on relative model weights), since alignment selection is determined by the probability of the model path.

In the following section, we discuss the way to use probability gain to compute stochastic alignment.

\section{Stochastic Alignment}\label{sec:method}
In this section, we introduce our stochastic alignment technique, which matches an observed trace to a stochastic process model.
The problem is solved by translating it to a shortest-path search problem that can be addressed by the a-star algorithm.
First, we describe the Pareto front that exists for the probability of model paths and their edit distance to the trace. 
Second, we introduce a loss function with a user-defined balance factor, which enables business analysts to weigh the trade-off and compute an alignment that satisfies the Pareto-optimality.
Third, we translate our problem to a shortest path problem using a standard a-star algorithm~\cite{DBLP:journals/tssc/HartNR68} and provide a heuristic for SLPNs.

A model path with a low edit distance to a given trace effectively explains where the model and the log trace disagree.
However, suppose such a model path has a marginal probability according to the stochastic model, then it may be deemed as less relevant to match the trace. 
This scenario illustrates that the probability of a selected model path and its edit distance to the trace are two potentially competing objectives when explaining deviations.
This trade-off naturally leads to a Pareto front, which we illustrate in \Cref{fig:pareto}.

\begin{table}[t]
\begin{minipage}[c]{0.5\linewidth}
\centering
\centering
    \includegraphics[width=5.5cm,height=3cm]{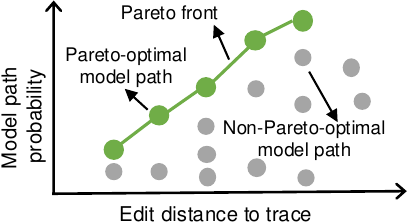}
    \captionof{figure}{A Pareto front for model paths.}
    \label{fig:pareto}
\end{minipage}
\hfill
\begin{minipage}[c]{0.48\linewidth}
\vspace{-0.4cm}
\caption{The probability of model paths in~\cref{fig:example_slpn}, and their edit distance to $\langle a,d,c\rangle$.
\label{tab:slpn_path}}
\begin{tabularx}{\linewidth}
{l@{\extracolsep{\stretch{1}}}*{3}{c}@{}}
\toprule
\makecell[c]{model\\path}  &\makecell[c]{model\\trace}& probability & \makecell[c]{edit\\distance} \\
\midrule
$\langle t_1,t_3\rangle$ & $\langle a,c\rangle$&\nicefrac{5}{500} & 1 \\
$\langle t_2,t_3,t_4\rangle$&$\langle b,c,d\rangle$ & \nicefrac{297}{500}& 4\\
$\langle t_2,t_4,t_3\rangle$& $\langle b,d,c\rangle$ &\nicefrac{198}{500} & 2\\
\bottomrule
    \end{tabularx}
\end{minipage}
\end{table}

In this work, a Pareto front consists of model paths representing an optimal trade-off: any reduction in edit distance would necessarily decrease the model path probability, and any increase in model path probability would necessarily increase its edit distance to the trace.
For instance, as illustrated in~\cref{tab:slpn_path}, given the observed trace $\langle a,d,c\rangle$, there are three Pareto-optimal model paths allowed by $N_2$, namely $\langle t_1,t_3\rangle$, $\langle t_2,t_3,t_4\rangle$, and $\langle t_2,t_4,t_3\rangle$.

\subsection{Loss Function}
Selecting an alignment from the Pareto front requires balancing two competing objectives: the model path probability $p$ and its edit distance $d$ to the trace. 
This can be quantified through a function $f_1 \colon d\rightarrow \mathbb{R}_{\ge 0}$ for edit distance $d$ and a function $f_2 \colon p\rightarrow \mathbb{R}_{\ge 0}$ for probability $p$.
As demonstrated previously, no single model path in \lmodel may be optimal for both $f_1$ and $f_2$ simultaneously. 
Thus, a stochastic alignment should be measured by a combined function $loss\colon (d,p)\rightarrow \mathbb{R}_{\ge 0}$ that addresses this trade-off, so that it satisfies the following properties.

\begin{property}
Given a trace, if two model paths have the same edit distance to it, the more likely model path yields a smaller value for the loss function. That is, let $d$ be the edit distance with $d>0$, for all $p_1 > p_2 \Rightarrow loss(d,p_1) < loss(d,p_2)$.\label{prop1}
\end{property}

\begin{property}
Given a trace, if two model paths are equally likely according to the stochastic model, the model path with a smaller edit distance to the trace yields a smaller value for the loss function. That is, let $p$ be the probability, for all $d_1 < d_2 \Rightarrow loss(d_1,p) < loss(d_2,p)$.
\end{property}

\begin{property}
Given a trace, the more likely model path with a smaller edit distance to the trace yields a smaller value for the loss function. That is, for all $0\le d_1 < d_2 \land p_1 >p_2 >0\Rightarrow loss(d_1,p) < loss(d_2,p)$.
\end{property}

To account for these properties, we propose a loss function to measure Pareto optimality as follows.
\begin{align}
   loss(d,p) = \begin{cases}
                (\lg (d+1))^{\alpha} & \text{ if } \alpha = 1\\
                (1 - \lg p)^{1-\alpha} & \text{ if } \alpha = 0\\
                (\lg (d+1))^{\alpha}\cdot (1 - \lg p)^{1-\alpha} & \text{ if } \alpha \in (0, 1)
            \end{cases}\label{def:loss}
\end{align}
In~\cref{def:loss}, $\lg (1+d)$ is base-10 logarithmic edit distance, in which we add one to avoid the negative-infinite $\lg 0$, and $(1-\lg p)$ represents the inverse of the logarithm of the path's probability.

The loss function is controlled by a user-defined balance factor $\alpha\in [0,1]$, which assigns a weight to the logarithmic probability function and logarithmic edit distance function, so business analysts can balance between these two perspectives.
It is trivial to see that the function adheres to Properties 1, 2, and 3 when $\alpha$ is not equal to 0 or 1.
For the special case of $\alpha = 1$, the model path with the least edit distance to the trace has the least loss. 
This corresponds to conventional alignment that only minimizes the deviations between the model path and the trace.
In another special case of $\alpha = 0$, the most likely model path has the least loss. 

\begin{figure}[t]
    \centering
    \begin{subfigure}[t]{0.34\textwidth}
        \centering
        \includegraphics[width=\linewidth]{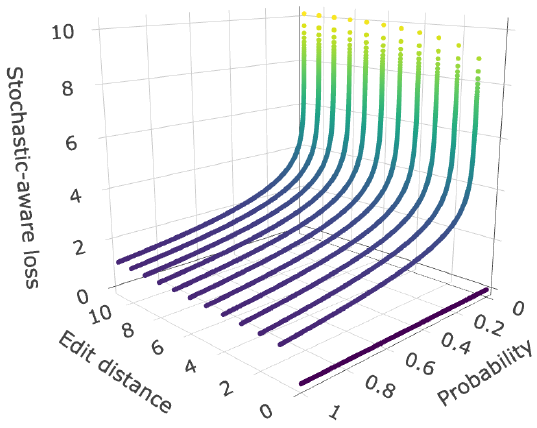}
        \captionsetup{font=small}
        \caption{$\alpha$=0.1.}
    \end{subfigure}%
    ~\hfill
    \begin{subfigure}[t]{0.31\textwidth}
        \centering
        \includegraphics[width=\linewidth]{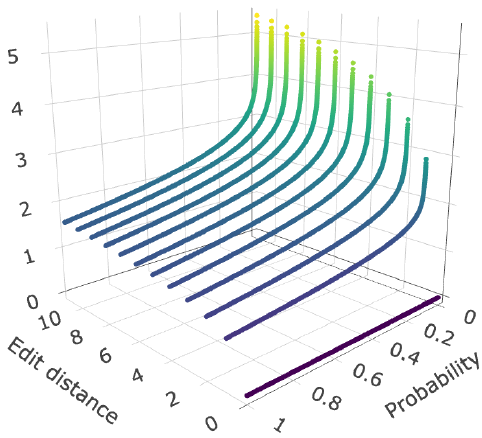}
        \captionsetup{font=small}
        \caption{$\alpha$=0.5.\label{fig:loss_05}}
    \end{subfigure}
    ~\hfill
    \begin{subfigure}[t]{0.31\textwidth}
        \centering
        \includegraphics[width=\linewidth]{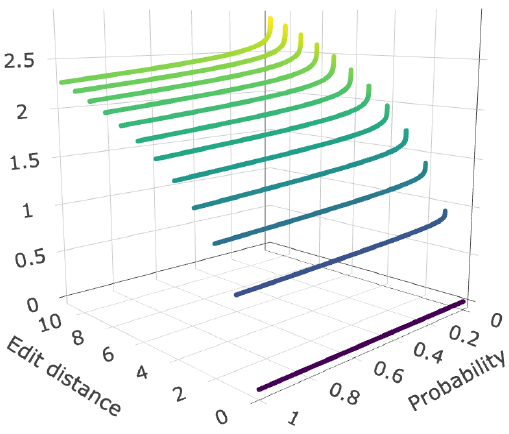}
        \captionsetup{font=small}
        \caption{$\alpha$=0.9.}
    \end{subfigure} 
    \caption{The value of loss function with different user-defined balance factor $\alpha$.\label{fig:loss}}
\end{figure} 

~\cref{fig:loss} illustrates the influence of $\alpha$.
The loss values for different $\alpha$ are not directly comparable, as a lower $\alpha$ leads to a generally higher loss for the same edit distance and model path probability.
If $\alpha$ is 0.5, the trade-off between the edit distance and the probability of the model path scales relatively evenly in both directions, as shown in~\cref{fig:loss_05}.
For example, a model path with a probability of 0.003 and an edit distance of 4 to the trace has a stochastic-aware loss smaller than a model path with a probability of 0.001 and an edit distance of 3.
When $\alpha$ is set to 0.1, the value of the loss function is based mainly on the probability of the path of the model, as a marginal increase in the probability would lead to a smaller loss from the probability range of 0 to 0.2.
For example, a model path with a probability of 0.0013 and an edit distance of 3 to the trace has a loss that is smaller than a model path with a probability of 0.001 and an edit distance of 2.
When $\alpha$ approaches 1, a model path with a small edit distance can hardly be overtaken by another model path with a larger edit distance.

~\cref{tab:different_alpha} presents the values of the loss function using different $\alpha$ for our running example. 
Different model paths are favored under different $\alpha$, as highlighted in bold. 
The example illustrates that, depending on the user-defined $\alpha$, the stochastic alignment computed changes accordingly.

\begin{table}[t]
  \centering
  \caption{The value of loss function for running example with different balance factors.}\label{tab:different_alpha}
  \small
\begin{tabularx}{\linewidth}
{c@{\extracolsep{\stretch{1}}}*{5}{c}@{}}
\toprule
path  & $\alpha{=}0$ & $\alpha{=}0.25$ &$\alpha{=}0.5$ &$\alpha{=}0.75$ &$\alpha{=}1$  \\
\midrule
 $\langle t_1,t_3\rangle$ & 3.000 & 1.688 & 0.950 & \textbf{0.535} & \textbf{0.301}  \\ 
 $\langle t_2,t_3,t_4\rangle$ & \textbf{1.226} & \textbf{1.065} & 0.926 & 0.804 & 0.699  \\ 
$\langle t_2,t_4,t_3\rangle$ & 1.402 & 1.071 & \textbf{0.818} & 0.625 & 0.477  \\ 
\bottomrule
    \end{tabularx}
\end{table}

\subsection{Computing a Stochastic Alignment}
Given a trace and an \lmodel of the process, we introduce a shortest path search to compute a stochastic alignment that minimizes the loss function.

\myparagraph{Applying a-star.} We solve our problem by identifying a sequence of moves in the synchronous product net that minimizes the loss function.
This optimization can be formulated as a shortest path problem~\cite{Adriansyah2014AligningOA}.
To guide the a-star search, we introduce a dual-component heuristic: one estimates the minimum remaining edit distance to reach a deadlock marking, while the other estimates the maximum remaining probability gain to a deadlock marking.
During search iteration, the $f(M)$ of a marking $M$ visited can be formulated as:
\begin{align}
      f(M) = \begin{cases}
                (\lg (g_d+h_d+1))^{\alpha} & \text{ if } \alpha = 1\\
                (1 - \lg (g_p*h_p))^{1-\alpha} & \text{ if } \alpha = 0\\
                (\lg (g_d+h_d+1))^{\alpha}\cdot (1 - \lg (g_p*h_p))^{1-\alpha} & \text{ if } \alpha \in (0, 1)
            \end{cases}
\end{align}
where $g_d$ and $g_p$ are the actual edit distance and probability gain from the initial marking to $M$, $h_d$ and $h_p$ are the estimated remaining edit distance and probability gain from $M$ to a deadlock marking. 
We first introduce the heuristic that underestimates the edit distance with Mixed-Integer Linear Programming (MILP).

\myparagraph{MILP-based Edit Distance Heuristic.} Let $N^\otimes= (P^\otimes,$ $T^\otimes,$ $F^\otimes,$ $A^\otimes,$ $\rho^\otimes$, $M_0^\otimes)$ be the synchronous product net of a trace and an \lmodel, with incidence matrix $\mathcal{I}$ and consumption matrix $\mathcal{C}$.
Then, we solve the following edit distance heuristic.
\begin{align}
    \text{minimize } & h_d =\vv{d}^{\mathsf{T}} \cdot \vv{x} \nonumber\\
    \text{such that } & \vv{x}(t) \in \mathbb{N} \land {} \label{def:cost_c1} \\
    & \vv{M_d} = \vv{M} + \mathcal{I} \cdot \vv{x} \land {} \label{def:cost_c2}\\
    & \forall_{t\in T^\otimes} \exists_{p\in P^\otimes} \ \vv{M_d}(p) < \mathcal{C}(p,t) \label{def:cost_c3}
\end{align}
In which $\vv{x}$ is a Parikh vector that describes the number of times each transition should be fired to reach a deadlock marking from $M$, and $\vv{d}$ is a $|N^\otimes.T|$-sized vector for the transition-based edit distance function.
The heuristic for edit distance $h_d$ is the result of the minimization.

Constraint~\eqref{def:cost_c1} specifies that each element in the solution vector $\vv{x}$ must be non-negative.
Constraints~\eqref{def:cost_c2} and~\eqref{def:cost_c3} specify that $M_d$ should be a deadlock marking that does not enable any transition.
Due to possibly (infinitely) many deadlock markings in an \lmodel, the synchronous net can also have multiple deadlock markings. 
Thus, the MILP-based edit distance heuristics differentiate from existing (I)LP-based heuristics~\cite{Adriansyah2014AligningOA,DBLP:conf/bpm/Dongen18} by not explicitly specifying a deadlock marking.
As no transition is enabled at $M_d$, at least one place in the preset of each transition does not contain enough tokens.
We use binary decision variables and apply the big-M method~\cite{DBLP:books/daglib/0022091} to construct constraints~\eqref{def:cost_c3}.
Therefore, the heuristic is computed by solving an MILP.

\begin{lemma}[$h_d$ provides a lower bound on the edit distance to the deadlock marking]\label{thm:h_d}
Let $N^\otimes= (P^\otimes, T^\otimes, F^\otimes, A^\otimes, \rho^\otimes, M_0^\otimes)$ be a synchronous product net, and $d:T^\otimes\rightarrow \mathcal{R}_{\ge 0}$ be an edit distance function. Given a marking $M\in \mathcal{B}(P^\otimes)$, and a firing sequence $\gamma \in {T^\otimes}^*$ such that $M[\gamma\rangle M_d$ and $T^\otimes_{M_d}=\emptyset$, it holds that $h_d \le d(\gamma)$.
\end{lemma}

This lemma follows from the fact that $h_d=\vv{d}^{\mathsf{T}}\cdot \vv{x}\le 
 d(\gamma)=\vv{d}^{\mathsf{T}}\cdot \vv{\gamma}$, otherwise $\vv{x}$ is not minimizing.
Likewise, we define another heuristic that overestimates the remaining probability of reaching the deadlock marking.

\myparagraph{MILP-based Probability Gain Heuristic.} Let $N^\otimes= (P^\otimes,$ $T^\otimes,$ $F^\otimes,$ $A^\otimes,$ $\rho^\otimes$, $M_0^\otimes)$ be the synchronous product net of a trace and an \lmodel, with incidence matrix $\mathcal{I}$ and consumption matrix $\mathcal{C}$.
Then, we solve the following probability gain heuristic.
\begin{align}
    \text{maximize } & h_p= \vv{p}^{\mathsf{T}} \cdot \vv{x} \label{def:lp_probability}\nonumber\\
    \text{such that } & \vv{x}(t)\in \mathbb{N} \land {} \\
    & \vv{M_d} = \vv{M} + \mathcal{I} \cdot \vv{x} \land {}\\
    & \forall_{t\in T^\otimes} \exists_{p\in P^\otimes} \ \vv{M_d}(p) < \mathcal{C}(p,t)
\end{align}
In which $\vv{x}$ is a Parikh vector, $\vv{p}$ is a $|N^\otimes.T|$-sized vector for the transition-based probability gain function, and $h_p$ is the maximized estimation of the remaining probability gain.

\begin{lemma}[$h_p$ provides an upper bound on the probability gain to the deadlock marking]\label{thm:bound}
Let $N^\otimes= (P^\otimes, T^\otimes, F^\otimes, A^\otimes, \rho^\otimes, M_0^\otimes)$ be a synchronous product net, and $p\colon T^\otimes\rightarrow \mathcal{R}_{\ge 0}$ be a probability gain function. Given a marking $M\in \mathcal{B}(P^\otimes)$, and a firing sequence $\gamma \in {T^{\otimes}}^*$ such that $M[\gamma\rangle M_d$ and $T^\otimes_{M_d}=\emptyset$, it holds that $h_p \ge p(\gamma)$.
\end{lemma}

This lemma follows from the fact that $h_p=\vv{p}^{\mathsf{T}}\cdot \vv{x}\ge p(\gamma)=\vv{d}^{\mathsf{T}}\cdot \vv{\gamma}$, otherwise $\vv{x}$ is not maximizing.
As the sub-heuristic for edit distance is minimizing and the sub-heuristic for probability gain is maximizing, the dual-heuristic for a-star is admissible.

\begin{lemma}[The dual-heuristic is admissible]\label{thm:admissible}
Let $N^\otimes= (P^\otimes, T^\otimes, F^\otimes, $$A^\otimes,$ $ \rho^\otimes, $$M_0^\otimes)$ be the synchronous product net, $g_d$ and $g_p$ be the actual edit distance and probability gain for the path from $M_0^\otimes$ to marking $M$, $h_d$ and $h_p$ be the heuristic of edit distance and probability gain for $M$. For all firing sequence $\gamma \in {T^\otimes}^*$ such that $M[\gamma\rangle M_d$ and $T^\otimes_{M_d}=\emptyset$, it holds that $h_d \le d(\gamma)$ and $h_p \ge p(\gamma)$. With the loss function~\cref{def:loss}, we have:
\begin{align}
\lg(g_d+h_d+1)^\alpha \le \lg(g_d+d(\gamma)+1)^\alpha, \text{ if } \alpha=0 \nonumber\\
(1-\lg (g_p*h_p))^{(1-\alpha)}\le (1-\lg (g_p*p(\gamma))^{(1-\alpha)}, \text{ if } \alpha =1\nonumber
\end{align}
The dual-heuristic for $\alpha\in(0,1)$ also lead to an underestimation of loss for $M$.
\end{lemma}

\myparagraph{Algorithmic description}
Using the heuristic for edit distance and probability gain, we propose a modified version of the a-star search algorithm.
The input of our algorithm is the synchronous product net of a trace and \lmodel, and the loss function defined by the user in~\cref{def:loss}. 
In each iteration, the marking with the minimized loss is chosen for expansion. 
If it is not a deadlock marking, the search expands over all subsequent markings.
The edit distance and the probability gain of the path when reaching a marking, combined with the heuristic, determine the loss for the expanded marking.
The algorithm terminates once a deadlock marking is found and returns an alignment.

\section{Evaluation}\label{sec:evaluation} 
Our technique for computing the stochastic alignment has been implemented and is publicly available.$\footnote{https://github.com/BPM-Research-Group/Ebi}$
We first examine the scalability of the technique using stochastic models mined from real-life event logs.
Subsequently, we conduct a case study to illustrate actionable insights gained from our approach.
All experiments were performed on a MacBook Pro, with an M2 Pro processor, 32 GB memory, and macOS Sequoia 15.
\subsection{Scalability} 
\begin{figure}[t]
    \centering
    \begin{subfigure}[t]{\textwidth}
        \centering
        \includegraphics[width=0.9\linewidth]{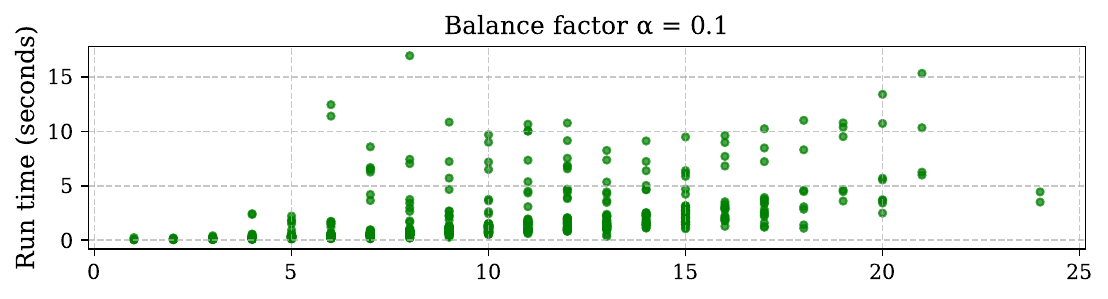}
    \end{subfigure}
        \begin{subfigure}[t]{\textwidth}
        \centering
        \includegraphics[width=0.9\linewidth]{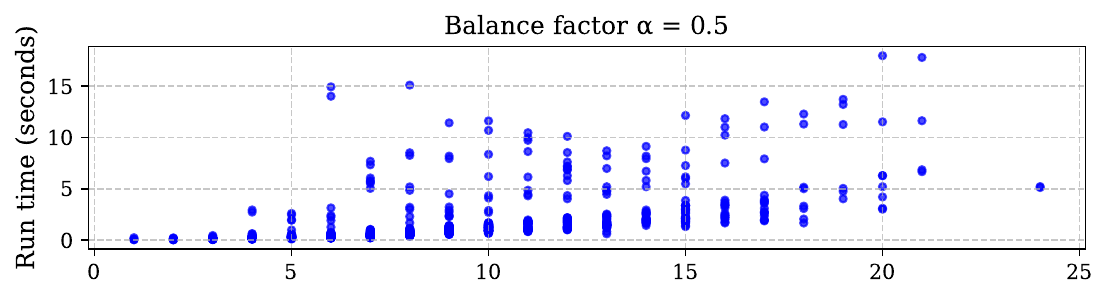}
    \end{subfigure}
        \begin{subfigure}[t]{\textwidth}
        \centering
        \includegraphics[width=0.9\linewidth]{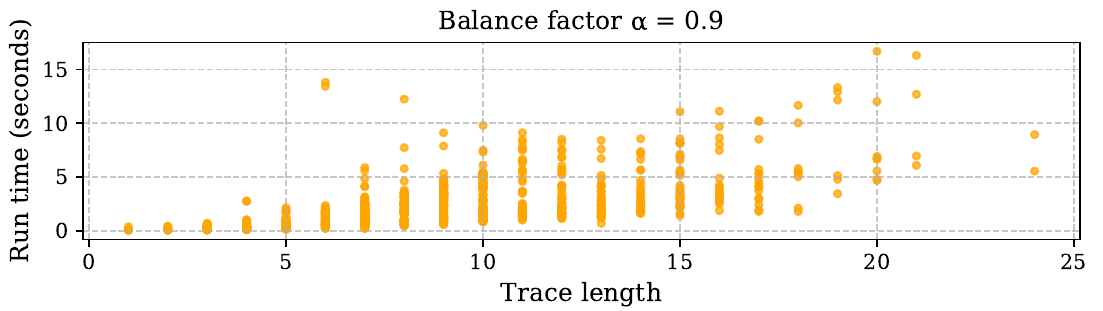}
    \end{subfigure}
    \caption{Running time versus trace length for different balance factors.\label{fig:experiment}}
\end{figure}

To evaluate the scalability of our technique, we used four event logs: road traffic fine management~\cite{road}, loan application~\cite{application}, payment request~\cite{bpi2020}, and domestic declaration~\cite{bpi2020}.
For each event log, we used Inductive Miner~\cite{DBLP:journals/kais/AugustoCDRP19} and Direct Follow Miner~\cite{DBLP:conf/icpm/LeemansPW19} to discover two control-flow models and then applied the alignment weight estimator~\cite{DBLP:conf/icpm/BurkeLW20} to obtain {\lmodel}s.
We then computed our stochastic alignment for each {\lmodel} and individual trace from 1040 model-trace pairs, for several values of $\alpha$ (0.1, 0.5, and 0.9), and measured the trace length and run time of our technique.
Each such run was repeated three times to reduce random effects, and the reported times are the averages over three runs.
\Cref{fig:experiment} shows the results.

Most alignments are computed in less than 10 seconds, with a few exceptions, though no computation took more than 20 seconds.
Our technique computes a shortest path in the reachability graph of the synchronous product net of the trace net and {\lmodel}.
The size of this reachability graph grows exponentially with the number of places in the product net, while the shortest path, without heuristics, can, in the worst case, be computed using Dijkstra's algorithm that relies on the Fibonacci heap priority queue in $O(E+V\mathit{log}V)$, where $V$ and $E$ are the numbers of vertices and edges in the reachability graph.
The few outliers in~\cref{fig:experiment} show scenarios where the a-star performs poorly, as the solution vector $\vv{x}$ for the MILP does not necessarily correspond to an actual path, and thus $\vv{x}$ needs to be recomputed often.
For different values of $\alpha$, we see a slight reduction in run time for lower $\alpha$, i.e., the model path probability having a larger influence, which may be due to alignments being computationally more expensive than finding the most likely model path.
In general, the results show that our technique is feasible for observed traces from real-life event logs and {\lmodel}s discovered from these logs.

\subsection{Applicability}
To evaluate the applicability of our approach, we applied our technique to analyze an observed trace from a travel permit process~\cite{bpi2020}. 
The process usually starts with a travel permit submitted by an employee.
After submission, the permit request is sent to the travel administration for approval. 
If approved, the request is forwarded to the budget owner and then to the supervisor. 
In some cases, the director also needs to approve the request.
The process ends with the payment being requested and handled. 

The experiment involves an auditor who has been equipped with the normative {\lmodel} and has selected a trace observed outside the training log, which stood out because it was nonconforming.
The goal is to analyze where the trace likely deviated from the model and to investigate whether any rules were violated. 

To obtain the normative stochastic process model, we split a training log consisting of the traces from 2018 onward.
Then, we apply Split Miner~\cite{DBLP:journals/kais/AugustoCDRP19} with default settings to discover a BPMN model, which is subsequently converted to an LPN, to which alignment weight estimation is applied~\cite{DBLP:conf/icpm/BurkeLW20}.

The observed trace selected by the auditor is $\langle$\textit{Permit submitted by employee, 
    Permit final approved by supervisor, 
    Start trip, 
    Declaration submitted by employee, 
    Declaration final approved by supervisor, 
    Request payment, 
    End trip, 
    Payment handled}$\rangle (\langle \mathit{PS},\mathit{PF},\mathit{ST},\mathit{DS},\mathit{DF},\mathit{RP},\mathit{ET},\mathit{PH}\rangle)$.
If the auditor only considers the edit distance, that is, applies our technique with $\alpha=1$ that outputs a conventional alignment, a closest model path is $m_1 = \langle$\textit{Permit submitted by employee, 
    Permit approved by administration, 
    Permit final approved by supervisor, 
    Start trip, 
    Declaration submitted by employee, 
    Declaration approved by administration, 
    Declaration final approved by supervisor, 
    Request payment, 
    Payment handled}$\rangle (\langle \mathit{PS},\mathit{PA},\mathit{PF},\mathit{ST},\mathit{DS},\mathit{DA},\mathit{DF},\mathit{RP},\mathit{PH}\rangle)$  with an edit distance of 3.
In particular, the model path $m_1$ contains two steps in which the administration is involved, while the trace does not contain these steps.
This may indicate a policy violation.
Moreover, this model path $m_1$ has a probability of $0.32\%$.

The auditor then applies our technique with $\alpha=0.5$ to obtain a model path that balances the likelihood of the model path and the edit distance.
This model path is $m_{0.5} = \langle$\textit{Permit submitted by employee, 
    Permit approved by administration, 
    Permit final approved by supervisor, 
    Start trip, 
    End trip, 
    Declaration submitted by employee, 
    Declaration approved by administration, 
    Declaration final approved by supervisor, 
    Request payment, 
    Payment handled}$\rangle(\langle \mathit{PS},\mathit{PA},\mathit{PF},\mathit{ST},\mathit{ET},\mathit{DS},\mathit{DA},$ $\mathit{DF},\mathit{RP},\mathit{PH}\rangle)$.
$m_{0.5}$ has a probability of 7.23$\%$ according to the model, which is 22 times higher than $m_1$.

\begin{figure}[t]
    \centering
    \includegraphics[width=\textwidth]{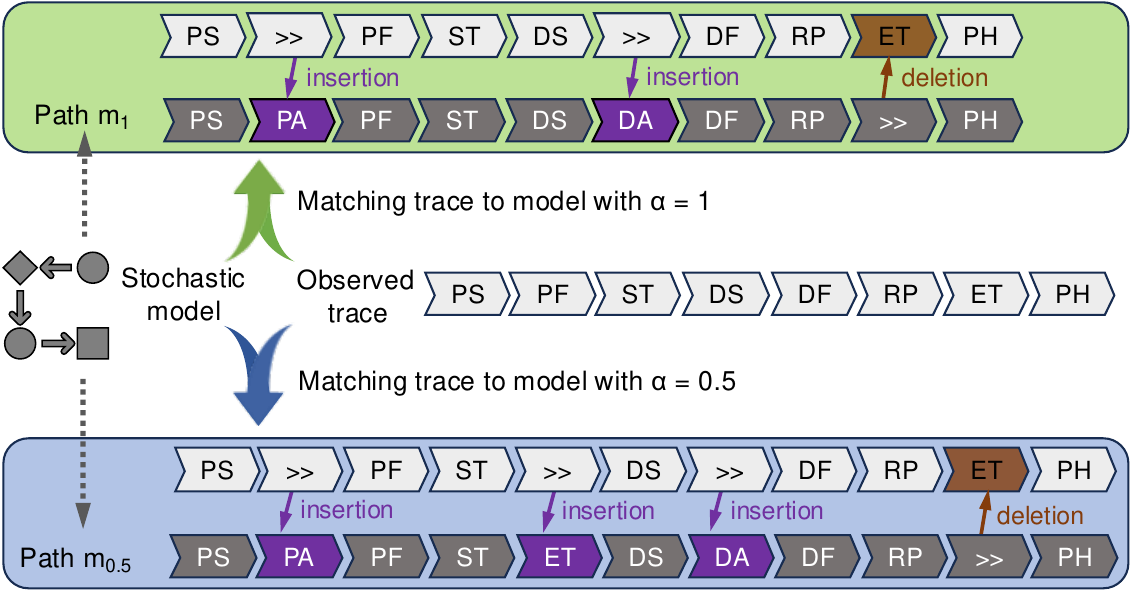}
    \caption{Matching the observed trace to the model with a different balance factor.}    \label{fig:evaluation_example}
\end{figure}

In~\cref{fig:evaluation_example}, we illustrate two alignment results returned to the auditor. 
The green region above corresponds to the conventional alignment with the least deviations, while the blue region below shows an alternative stochastic alignment that balances between deviation severity and path probability.
Compared to $m_1$, $m_{0.5}$ adds a deviation to the additional activity \textit{End trip} before \textit{Declaration submitted by the employee}.
As the model is normative, the auditor concludes that \textit{Permit approved by administration} and \textit{Declaration approved by administration} are missing from the observed trace (e.g., the execution of the activity was not performed following the formal procedure).
At the same time, the more likely model path suggests that the payment declaration should have been submitted after the end of the trip, rather than after the start of the trip.
The example shows that stochastic alignment generates new useful diagnostic insights for analysts in comparison to conventional alignment.

\section{Related Work\label{sec:related_work}}
Conformance checking provides mechanisms to relate modeled and observed process behavior.
The techniques in non-stochastic settings, such as token-replay~\cite{DBLP:journals/is/RozinatA08} and alignments~\cite{Adriansyah2014AligningOA,DBLP:conf/bpm/Dongen18,DBLP:conf/icpm/LiZ22}, have been extensively discussed.

The stochastic perspective of process behavior has enabled several new conformance checking techniques.
Entropic Relevance~\cite{DBLP:journals/is/AlkhammashPMG22} computes the average number of bits to compress the event log using the information of the likelihood of trace in a stochastic process model. 
Unit Earth Movers' Stochastic Conformance (uEMSC)~\cite{DBLP:conf/bpm/LeemansSA19} and Earth Movers' Stochastic Conformance (EMSC)~\cite{DBLP:journals/is/LeemansABP21} measure the effort to transform the distribution of traces in the log to the distribution described in the stochastic model. 
~\cite{DBLP:conf/icpm/RochaLA24} extends EMSC by considering partial trace mismatches, which is solved using Markovian abstraction~\cite{DBLP:journals/tkde/AugustoACDR22} for SLPNs.
Li et al.~\cite{li2024jensen} use the Jensen-Shannon Distance to compare two probability distributions over traces with their average probability distribution.
~\cite{DBLP:journals/is/IncertoVA25} applies the software performance engineering of Markovian processes from logs and compares it with model using uEMSC.
Although these techniques compute a numerical value between a stochastic model and an event log, the traces in the log are not explicitly matched to the model. 
Moreover, they are not applicable if an aggregated event log is not available.

Bogdanov et al.~\cite{DBLP:conf/bpm/BogdanovCG22} compute an alignment for a stochastically known trace with a probability function.
Alizadeh et al.~\cite{DBLP:conf/simpda/AlizadehLZ14a} extend the alignment cost functions by accounting for the probabilities of activities based on the process history, while Koorneef et al.~\cite{DBLP:conf/bpm/KoorneefSLR17} proposed to calculate the most probable alignment between a model and a log based on the probabilities of the observed process behavior in the event log.
However, the models used in these techniques are not stochastic. 
A work that uses stochastic models for alignment computation is the work by Bergami et al.~\cite{DBLP:conf/icpm/BergamiMMP21}, which matches a trace to a stochastic workflow net and returns a ranked list of approximated alignments.

The technique proposed in this paper is not directly comparable to existing techniques, as the input event data (individual trace vs. aggregated log~\cite{Adriansyah2014AligningOA,DBLP:conf/simpda/AlizadehLZ14a,DBLP:conf/bpm/KoorneefSLR17}) and models (stochastic vs. non-stochastic~\cite{Adriansyah2014AligningOA,DBLP:conf/bpm/KoorneefSLR17,DBLP:conf/bpm/BogdanovCG22}) differ fundamentally.
Furthermore, the output is one single alignment rather than a numerical value~\cite{DBLP:journals/is/AlkhammashPMG22,DBLP:journals/is/LeemansABP21,li2024jensen} or a list of alignments~\cite{DBLP:conf/icpm/BergamiMMP21}.

\section{Conclusion\label{sec:conclusion}}
In this paper, we propose a stochastic alignment technique that matches a trace with a likely path through a stochastic process model.
We introduce a user-defined parameter $\alpha$ that ranges from zero to one to allow for the trade-off between alignment cost and normative behavior.
For stochastic labeled Petri nets (SLPNs), we formalize this as an optimization problem and solve it by finding a shortest path in the synchronous product net using the a-star algorithm.
To navigate the search from the initial marking to potentially infinitely many deadlock markings, we use a dual-component heuristic.

Although the proposed stochastic alignment shares conceptual similarity with conventional alignment~\cite{Adriansyah2014AligningOA} by matching a trace to a model path, its stochastic perspective introduces additional constraints.
The first difference lies in how the technique handles stochastic models containing loops composed entirely of silent model moves. 
A stochastic alignment ($\alpha<1$) that identifies a path resulting from silent loops will incur a higher loss compared to one without such loops, as additional silent model moves represent more decision points, while the probability gains associated with silent model moves are constrained to be less than or equal to one.
Moreover, unless the balance factor is equal to 1, one cannot map all the traces to the most likely path of the model, as it is extremely unlikely that \emph{all} traces of the log took the most likely path and not a single trace took other less likely paths.
Enforcing all traces in a log to follow the ``most likely'' model paths would contradict real-world variability, as a subset of traces inherently follows less probable paths.

The technique has been implemented and is publicly available~\cite{DBLP:conf/icpm/LeemansLD24}. 
Our evaluation shows that the technique is feasible on real-life event data.
In a case study on a travel permit approval process, the technique identified deviations while quantifying the likelihood of the model path, offering auditors new and interesting insight into the conformance of the process. 

Accounting for multiple traces (or a complete event log) simultaneously is left for future work.
When matching a group of traces with a stochastic process model, instead of optimizing the match for each trace, one can search for a globally optimal match in which some likely model paths do not get overutilized by multiple traces using,
for instance, ideas similar to those applied in Earth Mover's Distance~\cite{DBLP:journals/is/LeemansABP21}.
We also plan to investigate a way to compute all Pareto-optimal model paths for a given trace.
Currently, the proposed technique is guided by a user-defined parameter to search for an alignment that identifies a Pareto-optimal model path.
We also plan to adopt other loss functions, such as the work in decision theory~\cite{DBLP:conf/ecai/DonadelloSHD24} that uses aggregation functions to balance multiple objectives.
\bibliography{bibliography} 
\bibliographystyle{splncs04} 
\end{document}